\documentclass[prb,amsmath,amssymb,twocolumn,showpacs,floatfix]{revtex4-1}
\usepackage{graphicx}
\usepackage[english]{babel}
\usepackage{times}
\usepackage{dcolumn}
\usepackage[usenames,dvipsnames]{color}
\usepackage{units}

\newcommand{\ps}{}
\definecolor{shadecolor}{gray}{.9}

\def \beq {\begin{equation}}
\def \edq {\end{equation}}
\def \bes {\begin{subequations}}
\def \eds {\end{subequations}}
\def \beqn {\begin{equation*}}
\def \edqn {\end{equation*}}

\def \dag {\dagger}

\def \up {\uparrow}
\def \down {\downarrow}

\def \sm {\sigma}
\def \bsm {\bar{\sigma}}

\def \veps {\varepsilon}
\def \weps {\widetilde{\varepsilon}}

\def \calh {{\cal{H}}}

\def \calg {{\cal{G}}}

\def \calf {{\cal{F}}}

\providecommand{\nbraket}[1]{\left\langle#1\right\rangle}
\providecommand{\dbraket}[1]{\langle\langle#1\rangle\rangle}
\begin{document}
\title{Non-equilibrium spin-current detection with a single Kondo impurity}
\author{Jong Soo Lim}
\affiliation{Institut de F\'{\i}sica Interdisciplin\`aria i de Sistemes Complexos
IFISC (CSIC-UIB), E-07122 Palma de Mallorca, Spain}
\affiliation{School of Physics, Korea Institute for Advanced Study, Seoul 130-722, Korea}
\author{Rosa L\'opez}
\affiliation{Institut de F\'{\i}sica Interdisciplin\`aria i de Sistemes Complexos
IFISC (CSIC-UIB), E-07122 Palma de Mallorca, Spain}
\affiliation{Departament de F\'{\i}sica,
Universitat de les Illes Balears, E-07122 Palma de Mallorca, Spain}
\author{Laurent Limot}
\affiliation{Institut de Physique et Chimie des Mat\'{e}riaux de Strasbourg$\text{,}$ Universit\'{e} de Strasbourg$\text{,}$ CNRS, 67034 Strasbourg, France}
\author{Pascal Simon}
\affiliation{Laboratoire de Physique des Solides, CNRS UMR-8502, Univ. Paris Sud, 91405 Orsay Cedex, France}
\date{\today}
\newcommand{\vc}[1]{{\mathbf{#1}}}
\newcommand{\vck}{\vc{k}}
\newcommand{\braket}[2]{\langle#1|#2\rangle}
\newcommand{\expv}[1]{\langle #1 \rangle}
\newcommand{\ket}[1]{| #1 \rangle}
\newcommand{\Tr}{\mathrm{Tr}} 
 \begin{abstract} 
We present a theoretical study based on the Anderson model of the transport properties of a Kondo impurity (atom or quantum dot) connected to ferromagnetic leads, which can sustain a non-equilibrium spin current. We analyze the case where the spin current is injected by an external source and when it is generated by the voltage bias. Due to the presence of ferromagnetic contacts, a static exchange field is produced that eventually destroys the Kondo correlations. We find that such a field can be compensated by an appropriated combination of the spin-dependent chemical potentials leading to the restoration of the Kondo resonance. In this respect, a Kondo impurity may be regarded as a very sensitive sensor for non-equilibrium spin phenomena. 
\end{abstract}
\pacs{72.10.Fk, 72.15.Qm,73.63.-b,68.37.Ef}
\maketitle

\section{Introduction}

In the last decades, there has been a revived interest in Kondo physics. This many-body effect is produced by high-order correlated tunneling events consisting of electronic spins hopping \textit{in} and \textit{out} a localized impurity, which ultimately lead to an efficient screening of the impurity spin. The Kondo effect has been extensively investigated for the anomalous behavior it produces on the resistivity versus temperature in bulk metals possessing magnetic impurities.~\cite{hewson1993} Experimental advances allow nowadays probing the Kondo effect in single objects through the detection of a zero-bias peak known as Kondo resonance. It is now possible to tackle non-trivial many-body effects in a controlled environment. The Kondo resonance has been investigated through scanning tunneling microscopy (STM) in single atoms either isolated~\cite{li98,madhavan98,pruser11} or coupled to other atoms,~\cite{manoharan00,wahl07,neel08,otte09} in single-atom contacts,~\cite{neel07,bork11,choi12} and in single molecules.~\cite{iancu06,fu07,torrente08,mugarza11,kawai12} It has also been successfully evidenced in nanoscale devices,~\cite{goldhabergordon1998a,goldhabergordon1998b,cronenwett1998,kouwenhoven2001,kouwenhoven1998} in particular quantum dots,~\cite{kouwenhoven1997,goldhabergordon1998a,goldhabergordon1998b, cronenwett1998,Schmid1998182,wiel2000} carbon nanotubes,~\cite{nygard2000,odom2000,jarillo} and nanowires.~\cite{PhysRevLett.107.076805}

Of particular interest\textemdash especially in the context of spintronics, is the issue of screening in the presence of a magnetic environment such as spin-polarized electrodes,~\cite{martinek2003b,PhysRevLett.90.116602,PhysRevLett.92.056601,pasupathy2004,simon2007,PhysRevB.83.155325,PhysRevLett.108.166605} and spin-polarized edge states.~\cite{PhysRevLett.96.046802} A spin-dependent hybridization for the spin-up and spin-down energy levels of the impurity is then predicted, resulting in an effective static magnetic field at the impurity site (this field can eventually be compensated by an external magnetic field).~\cite{PhysRevLett.92.056601,martinek2003b,PhysRevLett.108.166605} In the presence of ferromagnetism, the Kondo resonance therefore splits apart as confirmed experimentally.~\cite{pasupathy2004,fu12}

\begin{figure}
\centering
\includegraphics[width=0.4\textwidth]{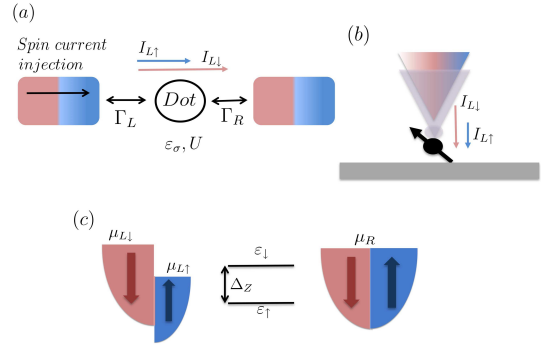}
\caption{\ps{Color on line.(a) Sketch of a quantum dot connected to a normal lead (right) and to a spin-accumulated lead (left). The spin current is injected by external means. The symbols $\Gamma_{L/R}$ represent the  hybridization between the dot level and the left/right leads. $I_{L,\uparrow}$ and $I_{L,\downarrow}$ denote the spin currents  for electrons with spin $\uparrow$ and $\downarrow$.(b) Schematic diagram of a magnetic atom adsorbed on a surface in contact with a spin-accumulated tip. (c) Spin accumulation gives rise to spin-dependent chemical potentials and eventually to a spin polarization. Both phenomena may lead to a splitting of the impurity levels. The color code is as follows: blue (resp. purple) denote electrons with spin $\uparrow$ (resp. $\downarrow$). }}
\label{fig:1}
\end{figure} 

While such a splitting is well understood, the impact of a non-equilibrium spin current on the Kondo resonance has so far been little addressed in correlated nanostructures. This remains an open question since a decade ago it was shown that a spin current flowing from a Co wire through a Cu(Fe) wire is able to strongly suppress the resistivity of the Cu(Fe) Kondo alloy near the interface.~\cite{PhysRevLett.90.016601} As demonstrated by Johnson~\cite{johnson} (see also Refs.~\onlinecite{valet-fert,johnson1,johnson2}), a spin current induces spin accumulation yielding spin-dependent chemical potentials $\mu_\up \ne \mu_{\down}$, which is equivalent to a spin bias. Using the equation of motion approach, Qi {\it et al.} analyzed the fate of a Kondo resonance in the presence of spin accumulation.~\cite{qi2008} They showed in particular that the Kondo resonance is split into two peaks attached to the two spin-dependent chemical potentials. \ps{ Kobayashi {\it et al.} ~\cite{kobayashi2010} recently validated this prediction by studying experimentally a Kondo quantum dot in contact with a spin accumulated electrode and two normal electrodes. A simplified geometry related to this experiment is sketched in Fig. 1a. }Besides demonstrating that the Kondo splitting can be controlled through spin accumulation, they also showed that the Kondo resonance may be restored through an external magnetic field.
 
Single-atom contacts with STM are another appealing way for investigating the interplay of a spin current with a Kondo impurity. In STM tunneling spectra, the Kondo resonance is detected as a Fano line shape due to the interference between electrons tunneling into the conduction band of the substrate and those involved into the Kondo state.~\cite{ujsaghy00,plihal01} When the tip is brought into contact with the atom such a picture remains valid although the Kondo resonance becomes more symmetric and of order the conductance unit $2e^2/h$.~\cite{choi12} As shown recently,~\cite{limot13} it it possible to introduce a spin current in the single-atom contact by using a ferromagnetic tip coated with a thick normal copper spacer (see Fig. 1b for a sketch of the setup). As in macroscopic spintronic devices,~\cite{johnson,johnson1,johnson2} the copper spacer aims at minimizing the direct or indirect magnetic exchange interactions between the cobalt atom and the tip. The Kondo splitting observed can then be assigned to spin accumulation in the copper spacer. In this respect, the Kondo resonance acts as a very sensitive local sensor for spin current. We want to emphasize that in the STM setup the tip is simultaneously the source of the spin accumulation and also the transport probe. Therefore the spin current becomes voltage-dependent contrary to Ref.~\onlinecite{kobayashi2010} where the spin current was supplied by an external spin-accumulated electrode while the differential conductance was probed using two different leads.

The purpose of this paper is to provide a microscopic description of a magnetic Kondo impurity embedded between a spin-polarized electrode able to carry a spin current and a normal metallic electrode. The impurity can be either artificial, such as a quantum dot (see Fig. 1a) or a genuine magnetic atom adsorbed on a surface (see Fig. 1b). We consider both cases in which the spin current is either driven by an external source (and therefore constant) or driven by the same electrode (and therefore voltage dependent). By using the equation of motion techniques~\cite{PhysRev.188.874,0305-4608-11-11-020,PhysRevLett.66.3048,PhysRevB.73.125338} and comparing various truncation methods to obtain a consistent picture, we show that the Kondo ground state depends sensitively on the spin polarization of the electrode and the spin accumulation that it generates. We investigate both the spin-resolved spectral density of the localized spin and the nonlinear conductance. In the case of a constant spin current, we demonstrate that spin accumulation leads to a splitting of the impurity Kondo resonance as shown previously\cite{qi2008,kobayashi2010} and schematically summarized in Fig. 1c. Taking into account both the static spin polarization and the spin accumulation, we show additionally that both effects can actually compensate each other and therefore the Kondo resonance can be restored. The case of a voltage dependent spin current turns out to be more subtle. Through a phenomenological approach we show that a non-linear dependence of the spin current with voltage bias is required to split the Kondo resonance. As we show in this work, our finite $Q$ approximation (which amounts to voltage-independent spin-dependent chemical potentials at large bias) turns out to be  rather accurate when the impurity is easy to spin polarize.
  
The plan of the paper is as follows: In Sec.~\ref{sec:model}, we introduce our model consisting of an impurity in contact with a spin-polarized electrode and a nomagnetic electrode. We also discuss
the method and approximation we use to tackle such a non-equilibrium interacting problem. In Sec. \ref{sec:q}, we study both analytically and numerically the case where one electrode has a finite spin polarization and sustains a constant spin accumulation. In Sec.~\ref{sec:q1}, we investigate the more subtle case where the spin accumulation becomes bias dependent. Finally, in Sec.~\ref{sec:conclu} we provide a summary of our main results and discuss some perspectives. Details on the truncated equation of motion approach used are presented in Appendix A.

\section{Model Hamiltonian and method}\label{sec:model}

We consider an impurity\textemdash a quantum dot or atom, coupled to left and right electrodes as depicted in Fig. 1.
We have in mind situations in which the quantum dot is used as a detector 
of a non-equilibrium spin accumulation, therefore we focus on the asymmetric situation in which
one electrode (the left one in Fig 1a or the STM tip in Fig 1b) may be partially polarized and able to sustain a spin current.

\ps{We assume that the impurity (the quantum dot or the magnetic adatom) correspond to a spin $S=1/2$ impurity.} 
In order to model the magnetic impurity, we consider an Anderson-type Hamiltonian
\begin{multline}
\calh = \sum_{\alpha,k,\sm} ({\veps_{\alpha k\sm}-\mu_{\alpha\sm}}) c_{\alpha k\sm}^{\dag} c_{\alpha k\sm} 
+ \sum_{\sm} \veps_{\sm} d_{\sm}^{\dag} d_{\sm} + U n_{\up} n_{\down}
\\  
+\sum_{\alpha,k,\sm} (V_{\alpha k\sm}c_{\alpha k\sm}^{\dag} d_{\sm} + h.c)\,.
\label{eq:ham}
\end{multline}
Here, $c_{\alpha k\sm}^{\dag} (c_{\alpha k\sm})$ denotes the creation (annihilation) operator in contact $\alpha$ and $d_{\sm}^{\dag} (d_{\sm})$ is the corresponding operator in the dot.
$V_{\alpha k\sm}$ describes a tunneling matrix element  between contacts and localized levels and can eventually be spin dependent. 
$U$ and $\varepsilon_\sigma$ parametrize the on-site Coulomb interaction and the spin-dependent localized energy level, respectively. 
Notice that an initial energy difference between localized levels, $\Delta_{Z}\neq \veps_\uparrow-\veps_\downarrow$,  may model an external magnetic field. 
As we emphasized in the introduction, a non-equilibrium spin accumulation entails spin-dependent chemical potentials $\mu_{\alpha\sm}$ and polarizations. 
The spin polarization in the contacts is lumped into spin dependent hybridization functions $\Gamma_{\alpha\sigma}$. 
Following Refs.~\onlinecite{PhysRevLett.90.116602,PhysRevLett.92.056601,martinek2003b}
we write a spin-polarization parameter as
\beq
P_\alpha = \frac{\Gamma_{\alpha\up}-\Gamma_{\alpha\down}}{\Gamma_{\alpha\up}+\Gamma_{\alpha\down}}\,,
\edq 
where $\Gamma_{\alpha\sm} = \Gamma_\alpha\left(1+ \sm P_{\alpha}\right)$ with $\Gamma_\alpha = \pi\sum_k|V_{\alpha k\sigma}|^2\rho_{\sigma}$.
$\rho_\sigma$ is the spin-dependent lead DOS at the Fermi energy, which is assumed flat for both electrodes.
Similarly, we introduce a spin bias parameter which is defined as
\begin{equation}
Q_{\alpha} = \frac{\mu_{\alpha\uparrow}-\mu_{\alpha\downarrow}}{\mu_{\alpha\uparrow} + \mu_{\alpha\downarrow}}\,.
\end{equation}

We consider the commonly used wideband limit for the tunneling rates where the hybridization $\Gamma_{\alpha\sm}$ are constant. 
Following Ref. [\onlinecite{meir1992}], the spin-dependent current $I_{\sm}$ reads
\beq\label{eq:current}
I_{\sm}\!\! =\!\!-\frac{4e}{h} \frac{\Gamma_{L\sm}\Gamma_{R\sm}}{\Gamma_{L\sm}+\Gamma_{R\sm}}\!\! \int d\omega~\left[f_{L\sm}(\omega) - f_{R\sm}(\omega)\right] \Im\left[G_{\sm,\sm}^r(\omega)\right] \,.
\edq
Here, $e$ is the elementary (positive) unit charge. 
and $ \Im$ denotes the imaginary part. Although the spin current expression may look simple, it is worth underlining that the retarded Green function $G^r$ is the exact bias-dependent Green function.
Computing such quantity remains a tremendous task. 
Since we are dealing with a non-equilibrium interacting problem, we must rely on some approximate approach able to capture qualitatively the physics.
We have employed the equation of motion technique to calculate the retarded Green function. 
This theoretical approach uses a truncated system of equations of motion for the retarded Green's function. 
There are several schemes for the truncation in order to obtain a close set of equations. 
In our case, we follow Refs.~\onlinecite{PhysRevLett.91.127203,0953-8984-19-34-346234} in order to compute the Keldysh Green functions. 
This procedure has been demonstrated to be suitable to treat systems with spin-polarized contacts. Details of the truncation scheme we have used can be found in the appendix A.

\section{Kondo resonance in the presence of a constant spin current and polarization}\label{sec:q}

In this section, we focus on a quantum dot connected to spin-polarized electrodes that are able to sustain a non-equilibrium spin current. In order to provide a qualitative understanding of the physics, it turns out to be useful to first perform a second order perturbation theory in the tunneling matrix elements, which generates an effective but non-equilibrium local
Zeeman term in the dot Hamiltonian.

\subsection{Effective magnetic fields}\label{sec:emf}
\ps{
Before presenting  the results of our numerical calculations, we would like to present an extension of the heuristic argument developed in [\onlinecite{Konig2005Lecture}] aiming at interpreting the effect of a finite polarization and/or finite spin current in a lead as an effective local exchange magnetic field viewed/felt by the spin impurity.
In order to calculate this effective magnetic field, we proceed as in [\onlinecite{Konig2005Lecture}] to investigate the functional dependence of the effective magnetic field on temperature and gate voltage.}
To do so, we derive an effective Hamiltonian $\mathcal{H}_\text{eff}$ using second-order perturbation theory. 
Physically, the split Kondo peak can be understood in terms of the dot valence instability (virtual charge fluctuation) and spin-dependent tunneling amplitudes. 
To deal with this instability, we perform a Schrieffer-Wolff-type transformation of the Hamiltonian given by Eq.~\eqref{eq:ham} and obtain
\begin{multline}
\mathcal{H}_{\text{spin}} = \sum_{\alpha,k}\sum_{\beta,q} \left[\frac{V_{\alpha k\up}V_{\beta q\up}}{\veps_{d\up}-\veps_{\beta q\up}} c_{\alpha k\up} c_{\beta q\up}^{\dag}
- \frac{V_{\alpha k\down}V_{\beta q\down}}{\veps_{d\down}-\veps_{\beta q\down}} c_{\alpha k\down} c_{\beta q\down}^{\dag}
\right.
\\
\left.
+ \frac{V_{\alpha k\up}V_{\beta q\up}}{U+\veps_{d\up}-\veps_{\beta q\up}} c_{\alpha k \up}^{\dag} c_{\beta q\up}
- \frac{V_{\alpha k\down}V_{\beta q\down}}{U+\veps_{d\down}-\veps_{\beta q\down}} c_{\alpha k\down}^{\dag} c_{\beta q\down}\right]S_z
\\ 
+\left[\cdots\right]\,,
\end{multline}
where $[\cdots]$ includes the usual terms corresponding to the spin-flip terms and potential scatterings that show up in the Kondo Hamiltonian.
At this point, unlike in the usual Schrieffer-Wolff transformation,
we employ a mean-field approximation for the lead electrons: $\langle c_{\alpha k\sm} c_{\beta q\sm}^{\dag}\rangle =[1-f(\veps_{\alpha k\sm})]\delta_{\alpha,\beta}\delta_{k,q}$, 
and $\langle c_{\beta q\sm}^{\dag} c_{\alpha k\sm}\rangle = f(\veps_{\alpha k\sm})\delta_{\alpha,\beta}\delta_{k,q}$. \cite{Konig2005Lecture}
The resulting effective Hamiltonian can be written as $\calh_{eff} = -B_{\text{eff}} S_z$.
The effective magnetic field generated by having spin-accumulation, i.e., spin-dependent chemical potentials and spin-polarized contacts then reads
\begin{multline}
B_{\text{eff}} \propto \sum_{\alpha} \int d\omega~ \left[ \frac{\Gamma_{\alpha\up} [1-f_{\alpha\up}(\omega)]}{\omega-\veps_{d\up}} - \frac{\Gamma_{\alpha\down} [1-f_{\alpha\down}(\omega)]}{\omega-\veps_{d\down}}
\right.
\\
\left.
+ \frac{\Gamma_{\alpha\up} f_{\alpha\up}(\omega)}{\omega-\veps_{d\up}-U} - \frac{\Gamma_{\alpha\down} f_{\alpha\down}(\omega)}{\omega-\veps_{d\down}-U} \right]\,,
\end{multline}
As emphasized in the introduction,  we will mainly focus on the asymmetrical situation and assume that the  spin accumulation and polarization occurs only in the left contact.
The general case can be trivially extended. We also assume spin-degenerate localized levels $\veps_{d\up}=\veps_{d\down}=\veps_d$. 
Therefore the spin-dependent chemical potentials are parametrized as follows:
\beq
\begin{split}
\mu_{L\up} &= \mu_L(1+Q),\,\,\text{and}\,\,\mu_{L\down}=\mu_L(1-Q) \,,
\\
\mu_{R\up} &= \mu_{R\down} = 0 \,,
\end{split}
\edq
and the lead polarization  $P$ is defined by:
\beq
\begin{split}
\Gamma_{L\up} &= \Gamma_L (1+P),\,\,\text{and}\,\,\Gamma_{L\down}=\Gamma_L(1-P) \,, \\
\Gamma_{R\sm} &= \Gamma_R \,.
\end{split}
\edq
With this parametrization, the effective field can be then written as
\begin{widetext}
\beq
B_{\text{eff}} \propto \Gamma_L \int d\omega~ \left[ \frac{(1+P) [1-f_{L\up}(\omega)]}{\omega-\veps_{d}} - \frac{(1-P) [1-f_{L\down}(\omega)]}{\omega-\veps_{d}} 
+ \frac{(1+P) f_{L\up}(\omega)}{\omega-\veps_{d}-U} - \frac{(1-P) f_{L\down}(\omega)}{\omega-\veps_{d}-U} \right].
\edq
Up to leading order in $P$ and $Q$, the previous expression simplifies to
\begin{multline}
B_{\text{eff}} \propto -2P\Gamma_L \Re\left\{ \Psi\left(\frac{1}{2}-i\beta\frac{(\veps_d-\mu_L)}{2\pi}\right) - \Psi\left(\frac{1}{2}-i\beta\frac{(\veps_d+U-\mu_L)}{2\pi}\right)\right\}
\\
+ 2Q\mu_L\Gamma_L \int d\omega~ f'(\omega-\mu_L)\left[\frac{1}{\omega-\veps_{d}} - \frac{1}{\omega-\veps_{d}-U} \right] \,,
\end{multline}
with $\beta = 1/k_B T$ and \ps{$\Psi$ denotes the digamma function defined as the logarithmic derivative of the gamma function}.
Taking the limit $T\to 0$, the effective magnetic field takes the compact form
\beq\label{eq:heff}
 B_{\text{eff}} \propto -2P\Gamma_L \ln \left|\frac{\veps_d-\mu_L}{\veps_d+U-\mu_L}\right| 
+2Q\mu_L\Gamma_L \frac{U}{(\veps_d-\mu_L)(\veps_d+U-\mu_L)}.
\edq
\end{widetext}
Note that for the particle-hole symmetry point ($|\veps_d-\mu_L|=U/2$) the effective field due to the finite spin polarization vanishes,
while  there still remains the effective field due to the spin-dependent chemical potentials.
Such effective magnetic field can be cancelled by applying an external static magnetic field $B_{\rm ext}$ such that $B_{\rm ext}+B_{\rm eff}=0$ as this has been shown experimentally by Kobayashi {\it et al.}.\cite{kobayashi2010}

However, we want to stress that this is not the only way to cancel this effective magnetic field. 
The two contributions related to the spin accumulation and static polarization may indeed have different sign. 
The sign of the former term is determined by the difference of the spin-dependent chemical potentials 
while the sign of the latter is fixed by the difference of the density of states at the Fermi energy between up and down electrons. 
Therefore this heuristic argument suggests that we can control the spin current independently of the polarization or vice-versa and thus restore the Kondo resonance. 
This could for example be achieved by controlling the polarization of the left lead. \ps{We check that this is indeed the case in the next subsection.}

\subsection{Spectral weights and differential conductance}

\begin{figure}
\centering
\includegraphics[width = 0.45\textwidth]{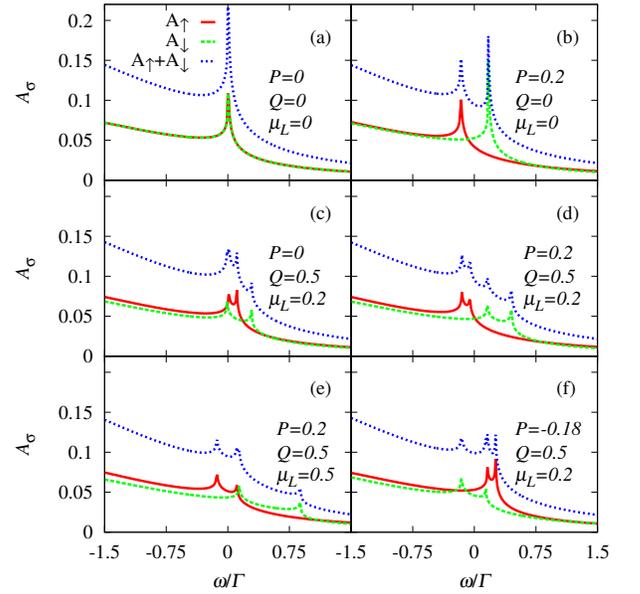}
\caption[Spectral weights vs $P$ and $Q$ ]
{Spectral weights vs $P$ and $Q$. Parameters: $\Gamma_L = \Gamma_R = 0.5$, $\veps_d = -3.5$, $D = 50$, $\mu_{R\sm} = 0$, $T = T_K$, and $U \to \infty$.}
\label{fig:2}
\end{figure}

We now present our results for the total and spin-resolved spectral weights which have been obtained with the truncated equation of motion approach. \cite{PhysRevLett.91.127203,0953-8984-19-34-346234} We work in units of $\Gamma_L + \Gamma_R=\Gamma=1$. 
We adopt  the following set of parameters  $\epsilon_d=-3.5$ and $D=50$. 
For simplicity, the $U\rightarrow \infty$ limit is considered, but our results can be generalized and remain qualitatively correct in the finite $U$ limit. 
Now, we vary $P$ and $Q$ and show the total and spin-resolved spectral density evolutions in Fig. 2. 
First, when both contacts are normal ($P=Q=0$), 
the low energy spectral density shows a single peak at the Fermi energy corresponding to the Kondo singularity [see Fig. \ref{fig:2}(a)]. 
When there is a finite polarization but no spin-dependent chemical potentials ($P\ne 0$, $Q=0$), 
the spin-$\up$ spectral density moves towards negative frequencies, whereas spin-$\down$ does the opposite,
resulting in a split Kondo resonance. This behavior is shown in Fig.~\ref{fig:2}(b). 
Next, we consider the situation with some degree of spin-dependent chemical potentials, but no spin polarization ($P=0$, $Q\ne 0$). 
Such a situation applies when a spin current is injected from an external terminal (see Ref. [\onlinecite{kobayashi2010}]).
We observe that the peaks in the spin-resolved spectral densities are located at $\omega_{\up(\down)} \approx \mu_{L\down(\up)}, \mu_R$; 
this is illustrated in Fig. \ref{fig:2}(c) for which we use $Q=0.5$, and $\mu_L=0.2$ leading to $\mu_{L\up}=0.3$, and 
$\mu_{L\down}=0.1$ while  $\mu_R=\mu_{R\up}=\mu_{R\down}=0$. 
\ps{ The  position of the peaks is determined by the poles of the impurity retarded green function (see Appendix A particularly Eq.~\eqref{eq:MeirIU}).
We find that the real part of the denominator of $\calg_{\sm,\sm}^r(\omega)$ has zeroes at $\mu_{\alpha\bsm}$ when the Kondo correlation develops.}
With a finite polarization, the renormalized levels become spin-dependent so that the peak positions are no longer at $\mu_{\alpha\sm}$ but depend on the degree of polarization $P$.
In general, when $P\ne 0$ and $Q\ne 0$, the total spectral density shows a four peak structure. 
However, if two of the four peaks encountered for $A_\uparrow$ and $A_\downarrow$ coincide, 
the total spectral function displays only three peaks. This situation is depicted in Fig. \ref{fig:2}(e) for which $P=0.2$, and $Q=0.5$ with $\mu_L=0.5$.
The three peaks can be also designed by considering $P=-0.18$ and $Q=0.5$ and $\mu_L=0.2$ as shown in Fig.~\ref{fig:2}(f).
Whereas the peak splitting in the spectral weights due to the static polarization can occur both under equilibrium and non equilibrium conditions, we want to stress again that the split spectral weights due to the spin-dependent chemical potentials can only occur under non-equilibrium conditions.


\begin{figure}
\centering
\includegraphics[width = 0.45\textwidth]{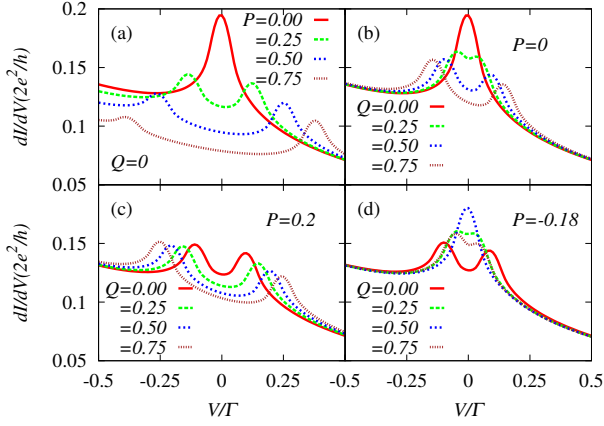}
\caption[Differential conductance vs $P$ and $Q$ ]
{Differential conductance vs $P$ and $Q$. 
Parameters: $\Gamma_{L\up}=0.3(1+P)$, $\Gamma_{L\down}=0.3(1-P)$, $\Gamma_{R\sm}=0.7$, $\mu_L = 0.2$, $\veps_d = -3.5$, $D = 50$, $T = 1.5T_K$, and $U \to \infty$.}
\label{fig:3}
\end{figure}

In addition, we investigate the nonlinear  conductance which is an experimentally accessible quantity. 
For practical purposes, it turns out to be more convenient to configure  the biases in such a way that the spin-dependent chemical potentials of the left contact are fixed, 
while the spin-independent chemical potential of the right contact, $\mu_{R\sm}=\mu_{L} + eV$, is varied.
The differential conductance reads:
\begin{multline}\label{eq:dif_cond}
\frac{dI_{\sm}}{dV} = \frac{4e^2}{h} \frac{\Gamma_{L\sm}\Gamma_{R\sm}}
{\Gamma_{L\sm}+\Gamma_{R\sm}}\int d\omega~ \left\{ \frac{d f_{R}(\omega)}{d(eV)} \Im\left[G^r_{\sm,\sm}(\omega)\right] \right. \\
\left. -[f_{L\sm}(\omega)-f_{R}(\omega)]\Im\left[\frac{dG^r_{\sm,\sm}(\omega)}{d(eV)}\right]\right\}.
\end{multline}

Figure \ref{fig:3}(a) illustrates the nonlinear conductance in the absence of spin-dependent chemical potentials ($Q=0$). 
The observed splitting is attributed to the effective field generated by the presence of the spin-polarized contacts.
Notice that the two peaks in the nonlinear conductance are almost symmetrically located and the splitting (therefore the peak positions) 
grows with $P$.
In the absence of spin polarization ($P=0$) but with spin-dependent chemical potentials, the nonlinear conductance shows similarly two peaks. 
Contrary to the $P=0$ case,  we easily identify that the $dI/dV$ conductance shows two peaks at $eV=\mu_{L\up(/\down)}-\mu_L$ [Fig. \ref{fig:3}(b)].
This corresponds to adjusting the spin-independent chemical potential of the right lead with the spin-dependent chemical potentials for the left lead.
This can be understood as follows. The leading part of the differential conductance is given by the first term in Eq.~\eqref{eq:dif_cond}.
At zero temperature, this term is proportional to the spin-dependent local DOS of the impurity ($A_{\sm}(\omega) = -\frac{1}{\pi}\Im\left[G^r_{\sm,\sm}(\omega)\right]$). 
When we neglect the second term in Eq.~(\ref{eq:dif_cond}), we therefore find that $dI_{\sm}/dV\propto A_{d\sm}(\mu_L+eV)$. 
The total differential conductance $dI/dV=\sum_{\sm} dI_{\sm}/dV$ will be maximum when the condition 
\beq
\label{eq:cond}
eV=\pm \frac{(\mu_{L\up}-\mu_{L\down})}{2} = \pm\mu_L Q~,
\edq
is satisfied.\cite{limot13}

In the presence of both spin bias and polarization ($Q\neq 0$, and $P\neq 0$), the nonlinear conductance for positive $P$ also exhibits two peaks 
and the splitting between two peaks widens compared with the $P=0$ cases.
On the contrary, for negative $P$, the splitting observed in the $dI/dV$ is reduced and eventually vanishes at some particular value of $Q$. 
This is shown in Fig.~\ref{fig:3}(d) where the nonlinear conductance shows a single resonance for $P=-0.18$ and $Q=0.5$. 
In this case, the Kondo effect is restored.  \ps{From the heuristic argument given in Sec. \ref{sec:emf}, we can interprete this restoration of the resonance as the compensation of the effective fields generated by the spin dependent polarization and the spin bias. }
Our numerical calculation therefore confirm nicely the qualitative results we discuss in Sec. \ref{sec:emf} that the accumulation spin current
and the static polarization may have antagonist effect on the Kondo resonance which results in its restoration.

\section{Kondo resonance in the presence of a bias-dependent spin accumulation}\label{sec:q1}

In the previous section, we studied the case where the spin current is injected by an external terminal as in Ref.~\onlinecite{kobayashi2010}. This was inherently a non-equilibrium situation even at $V=0$. Let us now consider the two-terminal situation with the left lead spin-polarized. We assume that at equilibrium (for $V=0$) no spin current is generated but only a static magnetic spin-polarization $P$. This implies $\mu_{L\up}=\mu_{L\down}$ at $V=0$. A finite $V$ generates both charge and spin currents. The purpose of this section is to analyze what will be the effect of such a bias-dependent spin current on the Kondo resonance. As we have seen, a finite spin current generates spin-dependent chemical potentials that are now encoded through the function $Q(V)$ verifying $Q(0)=0$. 

Let us first discuss the case where a static induced Zeeman field is not present, i.e., $B_{\rm eff}=0$.  Note that this can be achieved by fine tuning the gate voltage of a quantum dot to the particle-hole symmetric point according to Eq.~\eqref{eq:heff} where $B_{\rm eff}=0$. In the STM setup, this corresponds to the situation where $P=0$. This was implemented experimentally in Ref.~\onlinecite{limot13} by coating a magnetic tip with several layers of copper, in other words by introducing a normal metallic spacer between the spin-polarized tip and the atom. 

One may first try to expand the function $Q(V)$ in powers of $V$ which should correctly capture the behavior at low voltage. 
Keeping the first linear order in $V$ such that $Q(V)\approx a V$, we found numerically by solving the equation of motion for each value of the bias that the Kondo resonance does not split
(we take $P=0$). Indeed, this is consistent with the condition given in  Eq.~\eqref{eq:cond} which implies $a=0$. Moreover, at small bias, we expect a small non-equilibrium
effective magnetic field according to Eq.~\eqref{eq:heff} which does not split the Kondo resonance for $B_{\rm eff}\lesssim T_K$.
From  Eq.~\eqref{eq:current}, we see that the spin current $I_{\rm spin}=I_{L\up}-I_{L\down}$ is obviously a function of the bias $V$ but also of $Q(V)$. 
However, in the left lead, we expect $\Delta\mu_L=\mu_{L\up}-\mu_{L\down}=2\mu_L Q$ to be proportional to $I_{\rm spin}$.\cite{valet-fert}
Therefore, the function $Q(V)$ is highly non-linear and needs {\it a priori} to be determined self-consistently. 
However, this turns out to be a very difficult task (this is a non-linear and non-equilibrium interacting problem).  
We have chosen a different and simpler strategy by assuming various phenomenological forms for $Q(V)$ taking into account 
the constraints imposed on the function $Q(V)$  at small and large bias. 

\ps{Since all the  current must proceed through the single impurity states, the spectral  weight of the
impurity necessarily  limits the total amount of current. In other words,  a finite spin current entails a finite spin accumulation which cannot grow infinitely in a nanoscale structure. Therefore, the function $Q(V)$ is upper bounded and must converge asymptotically to a constant $Q_0$ at large $V$.}

\begin{figure}
\begin{center}
\includegraphics[width=0.5\textwidth]{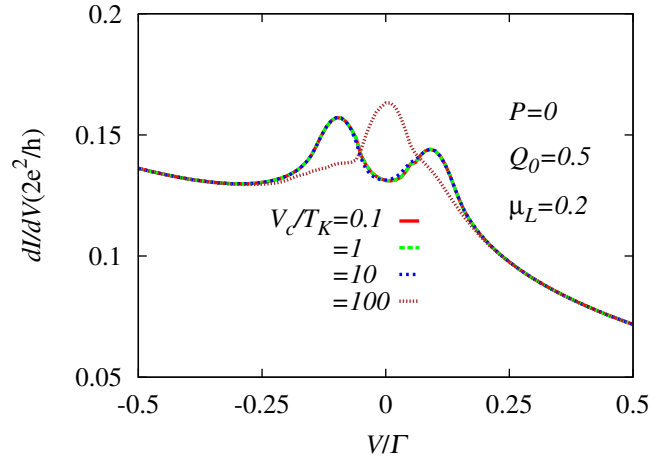}
\end{center}
\caption{Differential conductance for $P=0$ and in presence of bias dependent spin chemical potentials determined by $Q_1(V)=Q_0\tanh\left(-|V|/V_c\right)$ for different values of $V_c$. 
The other parameters are the same as before.}
\label{fig:bias1}
\end{figure}

In order to reconcile both the small and large $V$ limits we discussed, we have tested  the following phenomenological forms for $Q(V)$ given by
\bes
\begin{align}
Q_1(V) &= Q_0 \tanh\left(|V|/V_c\right)  \label{eq:q1v} \,, \\
Q_2(V) &= Q_0\left[1-\exp\left(-|V|/V_c\right)\right] \label{eq:q2v} \,.
\end{align}
\eds 
At large $|V|$, $Q_i(V)\to Q_0$, with $i=1,2$.
We have used $V_c$ as a phenomenological energy scale which is  related to the impurity orbital non-equilibrium polarization.
The larger $V_c$, the less susceptible to be spin polarized the impurity is. 
When $V_c$ is small compared to the voltage bias range explored, one can neglect the exponential and $Q_{1/2}(V)\approx Q_0$. We therefore recover the constant $Q$ case studied in Sec. \ref{sec:q}, which as we have seen leads to a splitting of the Kondo resonance.
Since the main energy scale entering into our problem is  the bare Kondo temperature $T_K$, one has to compare $V_c$ to $T_K$. 
\ps{Note here, that the bare Kondo temperature is the Kondo temperature the impurity would acquire in absence of polarization or spin accumulation.}
We have first  computed the differential conductance for $Q_0=0.5$ without any static polarization ($P=0$) for different values of $V_c$ using the function $Q_1(V)$. The results are summarized in 
Fig. \ref{fig:bias1}. We found that for $V_c\lesssim 10 T_K$, the constant-$Q$ approximation provides  results qualitatively similar to the constant-$Q$ approximation with a peak splitting. 
Only for large value of $V_c\gg 10T_K$, do we recover a Kondo resonance. Indeed for large $V_c\gg 10T_K$, we can expand the functions $Q_{1/2}(V)$ in $V/V_c$ since we are interested in bias of order of a few $T_K$. 
One can check that keeping the lowest terms of the expansion does not lead to a splitting of the peak using Eq.~\eqref{eq:cond}. We have also performed the same calculations with the function $Q_2(V)$ defined in Eq.~\eqref{eq:q2v}. 
\ps{The results  are almost  similar to the ones in Fig. \ref{fig:bias1}.}

\begin{figure}
\begin{center}
\includegraphics[width=0.5\textwidth,clip=]{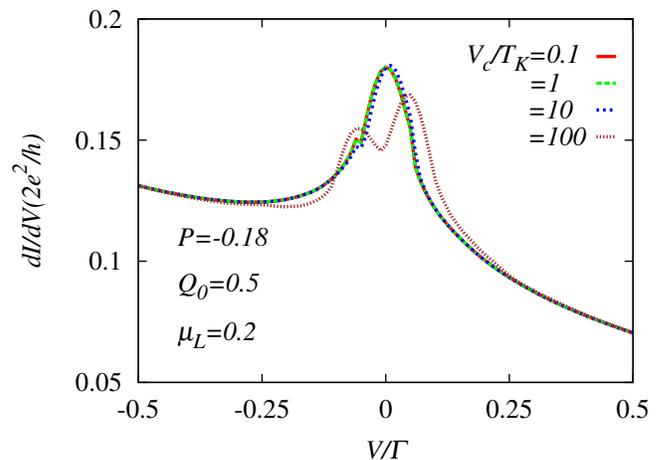}
\end{center}
\caption{Differential conductance for  $P=-0.18$  and $Q_2(V)=Q_0\left[1-\exp\left(|V|/V_c\right)\right]$ for different values of $V_c$. 
The other parameters are the same as before.}
\label{fig:bias2}
\end{figure}

We have also considered the case of a finite polarization $P$. When $P$ and $Q$ have the same sign, the Kondo resonance is always split as explained in Sec. \ref{sec:q}. We have computed
the differential conductance for $P=-0.18$ and $Q_0=0.5$ for different  values of $V_c$ using $Q_2(V)$. Our results are shown in 
Fig. \ref{fig:bias2}b. We found that for $V_c\lesssim 10 T_K$, as in the constant Q approximation, 
the static polarization and the spin bias-dependent chemical potential provide opposite effects which result in a restoration of the Kondo peak. 
Only for large value of $V_c\gg 10T_K$, are we dominated by the static polarization, which entails a splitting of the Kondo peak. 

\section{Conclusion}\label{sec:conclu}
In this paper we have studied a ferromagnetic-impurity-normal geometry paying attention to the fact that the ferromagnetic electrode is able to also sustain a spin current. 
Such a generic geometry can describe an artificial impurity (a quantum dot, a molecule, a carbon nanotube, etc.) or a genuine impurity contacted between a spin-polarized electrode and a normal one. For STM experiments, this corresponds to a magnetic adatom adsorbed on a metallic surface and in contact with a spin-polarized tip.
We have first considered the case in which the spin current is injected in one electrode and therefore maintained constant as realized experimentally in Ref.~\onlinecite{kobayashi2010}.
By computing the spectral functions and the differential conductance, we found that this leads to a splitting of the Kondo resonance. In that respect, the Kondo effect turns out to be a very sensitive phenomenon to detect a microscopic non-equilibrium spin current. We also studied the case where the spin current is injected into a spin-polarized electrode. Since the non-equilibrium spin accumulation and the equilibrium polarization have different origin, they can have antagonist effects. We have shown both analytically and numerically that the Kondo resonance can be restored when this is the case. Since the effect of the static polarization on an artificial impurity can be controlled by a gate voltage, this offers a knob to observe such a restoration of the Kondo resonance.
We have also considered the case where a spin current is generated when a voltage is applied between the two electrodes. Under this hypothesis, the spin-dependent chemical potentials become voltage dependent. Assuming simple phenomenological functions for this dependence that match the low-voltage and large-voltage case, we were able to show that the splitting of the Kondo resonance depends on a energy scale $V_c$ which can be related to  the polarisability of the impurity.  For $V_c/T_K\gg 1$, no splitting is found which corresponds to the case where the generated spin current at a fixed voltage $V$ is too small to split the resonance of width of order $T_K$. In the other limit, we recover results similar to the constant spin accumulation. We think that this semi-phenomenological treatment captures the essential physics. Future work will be necessary to describe this phenomenon microscopically without further assumptions. 

We have analyzed in the paper an anisotropic situation in which only one electrode is spin polarized. A natural extension of the present analysis is the more isotropic case in which the magnetic impurity is coupled to two spin-polarized electrodes that are able to sustain a spin current. Such a situation may apply to atomic contacts made from ferromagnetic materials where the observation of Kondo-Fano line shapes in the conductance have been reported.\cite{calvo09,calvo12}
  
\begin{acknowledgements}
We would like to thank M.V. Rastei for fruitful discussions. J.-S.L. and R.L. were supported by MICINN Grant No. FIS2011-2352. P.S. has benefited from financial support from the ANR under Contract No. DYMESYS (ANR 2011-IS04-001-01).
\end{acknowledgements}

\appendix

\section{Explicit expressions for the retarded Green's function $\calg_{\sm,\sm}^r(\omega)$}\label{appA}

In this appendix we derive the retarded Green's function to be used for the calculation of the differential conductance. For that purpose we
employ the equation of motion technique \cite{PhysRev.188.874,0305-4608-11-11-020,PhysRevLett.66.3048,PhysRevB.73.125338} and in particular the truncation schemes proposed in Refs.~\onlinecite{PhysRevLett.91.127203,0953-8984-19-34-346234}. These truncation procedures have been demonstrated to
describe properly systems attached to spin-polarized contacts.  The retarded impurity Green function is defined as
\beq
\calg_{\sm,\sm}^r(\omega) \equiv \dbraket{d_{\sm},d_{\sm}^{\dag}}_{\omega}^r = \int dt~ e^{i\omega t} \dbraket{d_{\sm},d_{\sm}^{\dag}}_{t}^r\,,
\edq
where
\beq
\dbraket{d_{\sm},d_{\sm}^{\dag}}_{t}^r = -i\Theta(t) \nbraket{[d_{\sm}(t),d_{\sm}^{\dag}(0)]}\,.
\edq
For a generic two particle operator  $\dbraket{A,B}_{\omega}^r$ we have for its equation of motion
\beq
\omega\dbraket{A,B}_{\omega}^r + \dbraket{[\calh,A],B}_{\omega}^r = \nbraket{[A,B]_+}\,.
\edq
When $A=d_\sigma$, $B=d^\dagger_\sigma$ the previous expression gives rise the equation of motion for the impurity Green function in the frequency domain.  
The imaginary part $i0^+$  going alongside $\omega$ is implicitly assumed. 
To simplify the notations, hereafter we write the retarded Green's functions $\dbraket{A,d_{\sm}^{\dag}}_{\omega}^r$ as $\dbraket{A}$.
The first equations of motion in the hierarchy are 
\begin{multline}
(\omega-\veps_{\sm})\dbraket{d_{\sm}} = 1\!\! +\!\! U\dbraket{d_{\sm}n_{\bsm}} + \sum_{\alpha,k} V_{\alpha k\sm} \dbraket{c_{\alpha k\sm}}\,,\\
(\omega-\veps_{\alpha k\sm})\dbraket{c_{\alpha k\sm}} =\!\! V_{\alpha k\sm} \dbraket{d_{\sm}}\,,
\end{multline}
where we take $V_{\alpha k\sm}$ as real.
Then, we have
\beq\label{eq:dret}
\left(\omega-\veps_{\sm} - \Sigma_{0\sm}(\omega)\right)\dbraket{d_{\sm}} = 1 + U\dbraket{d_{\sm}n_{\bsm}}\,,
\edq
where we define $\Sigma_{0\sm}(\omega)$ as the hopping selfenergy 
\beq
\Sigma_{0\sm}(\omega) = \sum_{\alpha,k} \frac{|V_{\alpha k\sm}|^2}{\omega-\veps_{\alpha k\sm}} \,.
\edq
To go to the next order we need to calculate $\dbraket{d_{\sm}n_{\bsm}}$ as well:
\begin{multline}
(\omega-\veps_{\sm}-U) \dbraket{d_{\sm}n_{\bsm}} = \nbraket{n_{\bsm}} + \sum_{\alpha,k} V_{\alpha k\sm} \dbraket{c_{\alpha k\sm}n_{\bsm}} \\ 
-\sum_{\alpha,k} V_{\alpha k\bsm} \dbraket{c_{\alpha k\bsm}^{\dag} d_{\bsm} d_{\sm}} + \sum_{\alpha,k} V_{\alpha k\bsm} \dbraket{d_{\bsm}^{\dag}c_{\alpha k\bsm}d_{\sm}} \,.
\label{eq:HI}
\end{multline}
The next step is to consider the equation of motion for each of the three higher order Green's functions  $\dbraket{c_{\alpha k\sm}n_{\bsm}} $, $\dbraket{c_{\alpha k\bsm}^{\dag} d_{\bsm} d_{\sm}} $, and $\dbraket{d_{\bsm}^{\dag}c_{\alpha k\bsm}d_{\sm}} $ that appear on the r.h.s. of Eq.~\eqref{eq:HI}. Below we write these three equation of motions approximated as
\begin{widetext}
\begin{eqnarray}
&&(\omega-\veps_{\alpha k\sm})\dbraket{c_{\alpha k\sm}n_{\bsm}} = V_{\alpha k\sm} \dbraket{d_{\sm}n_{\bsm}}
- \sum_{\beta,q} V_{\beta q\bsm} \dbraket{c_{\alpha k\sm} c_{\beta q\bsm}^{\dag}d_{\bsm}}
+ \sum_{\beta,q} V_{\beta q\bsm} \dbraket{c_{\alpha k\sm} d_{\bsm}^{\dag} c_{\beta q\bsm}}\,,
\label{eq:csnbs}
\\
&&(\omega-\veps_{\alpha k\bsm}+\veps_{\bsm}-\veps_{\sm}) \dbraket{d_{\bsm}^{\dag}c_{\alpha k\bsm} d_{\sm}} = \nbraket{d_{\bsm}^{\dag}c_{\alpha k\bsm}} + V_{\alpha k\bsm}\dbraket{d_{\sm}n_{\bsm}} 
\label{eq:dcd}  
\nonumber \\
&-& \sum_{\beta,q} V_{\beta q\bsm}\dbraket{c_{\beta q\bsm}^{\dag}c_{\alpha k\bsm}d_{\sm}} + \sum_{\beta,q} V_{\beta q\sm} \dbraket{d_{\bsm}^{\dag} c_{\alpha k\bsm} c_{\beta q\sm}}\,,
\\
&& \label{eq:threecor} \left(\omega + \veps_{\alpha k\bsm}-\veps_{\sm}-\veps_{\bsm} - U\right) \dbraket{c_{\alpha k\bsm}^{\dag}d_{\bsm}d_{\sm}}
=  \nbraket{c_{\alpha k\bsm}^{\dag}d_{\bsm}} - V_{\alpha k\bsm} \dbraket{d_{\sm}n_{\bsm}} 
\nonumber \\
&+& \sum_{\beta,q} V_{\beta q\sm}\dbraket{c_{\alpha k\bsm}^{\dag} d_{\bsm} c_{\beta q\sm}} + \sum_{\beta,q}V_{\beta q\bsm} \dbraket{c_{\alpha k\bsm}^{\dag}c_{\beta q\bsm} d_{\sm}}\,,
\end{eqnarray}
In the next step we truncate the system of equations keeping the Kondo correlations. To abbreviate the notation, we introduce a shorthand $\calf_{a;b}^{\sm} \equiv \nbraket{c_{a\sm}^{\dag}c_{b\sm}}$. We consider the onset of Kondo correlations by approximating Eqs.~\eqref{eq:csnbs}, \eqref{eq:threecor},  and  \eqref{eq:dcd} in the following way:
\begin{eqnarray}\label{sel1}
&&\sum_{\alpha,k} V_{\alpha k\sm} \dbraket{c_{\alpha k\sm}n_{\bsm}} \approx \sum_{\alpha,k} \frac{|V_{\alpha k\sm}|^2}{\omega-\veps_{\alpha k\sm}} \dbraket{d_{\sm}n_{\bsm}}
= \Sigma_{0\sm}(\omega) \dbraket{d_{\sm}n_{\bsm}} \,, 
\\
&&\sum_{\alpha,k} V_{\alpha k\bsm} \dbraket{d_{\bsm}^{\dag}c_{\alpha k\bsm} d_{\sm}}\approx \Sigma_{0\bsm}(\omega,\bsm;\sm,\alpha k\bsm)\dbraket{d_{\sm}n_{\bsm}}\, \nonumber \\ 
&+& \sum_{\alpha,k} \frac{V_{\alpha k\bsm}\nbraket{d_{\bsm}^{\dag}c_{\alpha k\bsm}}}{\omega-\veps_{\alpha k\bsm}+\veps_{\bsm}-\veps_{\sm}} \left(1 + \Sigma_{0\sm}(\omega) \dbraket{d_{\sm}}\right)
- \sum_{\alpha,k}\sum_{\beta,q} \frac{V_{\beta q\bsm}V_{\alpha k\bsm}\calf_{\beta q;\alpha k}^{\bsm}}{\omega-\veps_{\alpha k\bsm}+\veps_{\bsm}-\veps_{\sm}} \dbraket{d_{\sm}}\,,
\label{sel2}
\\
&&\sum_{\alpha,k} V_{\alpha k\bsm} \dbraket{c_{\alpha k\bsm}^{\dag}d_{\bsm}d_{\sm}} 
\approx  -\Sigma_{0\bsm}(\omega,\alpha k\bsm;\sm,\bsm,U) \dbraket{d_{\sm}n_{\bsm}} \nonumber \\
&+& \sum_{\alpha,k} \frac{V_{\alpha k\bsm}\nbraket{c_{\alpha k\bsm}^{\dag}d_{\bsm}}}{\omega+\veps_{\alpha k\bsm}-\veps_{\sm}-\veps_{\bsm}-U} \left(1 + \Sigma_{0\sm}(\omega)\dbraket{d_{\sm}}\right)
+ \sum_{\alpha,k}\sum_{\beta,q} \frac{V_{\alpha k\bsm}V_{\beta q\bsm}\calf_{\alpha k;\beta q}^{\bsm}}{\omega+\veps_{\alpha k\bsm}-\veps_{\sm}-\veps_{\bsm}-U} \dbraket{d_{\sm}}\,,
\label{sel3}
\end{eqnarray}
\end{widetext}
where the self-energies appearing in Eqs. \eqref{sel1}, \eqref{sel2}, and \eqref{sel3} are defined accordingly
\bes
\begin{align}
\Sigma_{0\bsm}(\omega,\bsm;\sm,\alpha k\bsm) &= \sum_{\alpha,k} \frac{|V_{\alpha k\bsm}|^2}{\omega-\veps_{\alpha k\bsm} + \veps_{\bsm}-\veps_{\sm}} \,,\\
\Sigma_{0\bsm}(\omega,\alpha k\bsm;\sm,\bsm,U) &= \sum_{\alpha,k} \frac{|V_{\alpha k\bsm}|^2}{\omega+\veps_{\alpha k\bsm}-\veps_{\sm}-\veps_{\bsm}-U}\,.
\end{align}
\eds
 Using the previous truncated high order propagators  we can show that Eq. \eqref{eq:HI} takes the expression
\beq
\dbraket{d_{\sm}n_{\bsm}} = \frac{\overline{\nbraket{n_{\bsm}}} - \Sigma_{2\sm}(\omega)\dbraket{d_{\sm}}}{\omega-\veps_{\sm}-U-\Sigma_{0\sm}(\omega)-\Sigma_{1\sm}(\omega)}\,,
\edq
where we have defined
\begin{eqnarray}
&&\overline{\nbraket{n_{\bsm}}} = \nbraket{n_{\bsm}}
+ \sum_{\alpha,k} \frac{V_{\alpha k\bsm}\nbraket{d_{\bsm}^{\dag}c_{\alpha k\bsm}}}{\omega-\veps_{\alpha k\bsm}+\veps_{\bsm}-\veps_{\sm}}
\nonumber \\
&-& \sum_{\alpha,k} \frac{V_{\alpha k\bsm}\nbraket{c_{\alpha k\bsm}^{\dag}d_{\bsm}}}{\omega+\veps_{\alpha k\bsm}-\veps_{\sm}-\veps_{\bsm}-U}\,,
\end{eqnarray}
\begin{widetext}
 together with the following self-energies
\begin{eqnarray}
\Sigma_{1\sm}(\omega) &=& \Sigma_{0\bsm}(\omega,\bsm;\sm,\alpha k\bsm) + \Sigma_{0\bsm}(\omega,\alpha k\bsm;\sm,\bsm,U)\,,
\\
\Sigma_{2\sm}(\omega) &=&
\sum_{\alpha,k}\sum_{\beta,q} \frac{V_{\beta q\bsm}V_{\alpha k\bsm}\calf_{\beta q;\alpha k}^{\bsm}}{\omega-\veps_{\alpha k\bsm} + \veps_{\bsm}-\veps_{\sm}}
+ \sum_{\alpha,k}\sum_{\beta,q} \frac{V_{\alpha k\bsm}V_{\beta q\bsm}\calf_{\alpha k;\beta q}^{\bsm}}{\omega+\veps_{k\bsm}-\veps_{\sm}-\veps_{\bsm}-U}
\nonumber \\
&-&\left[\sum_{\alpha,k} \frac{V_{\alpha k\bsm}\nbraket{d_{\bsm}^{\dag}c_{\alpha k\bsm}}}{\omega -\veps_{\alpha k\bsm} + \veps_{\bsm}-\veps_{\sm}}
- \sum_{\alpha,k} \frac{V_{\alpha k\bsm}\nbraket{c_{\alpha k\bsm}^{\dag}d_{\bsm}}}{\omega+\veps_{k\bsm}-\veps_{\sm}-\veps_{\bsm}-U} \right]\Sigma_{0\sm}(\omega)\,,
\end{eqnarray}
Taking back Eq. \eqref{eq:dret} into Eq. \eqref{eq:HI}  the dot retarded Green's function is finally obtained
\beq
\dbraket{d_{\sm}} = \frac{1 - \overline{\nbraket{n_{\bsm}}}}{\omega-\veps_{\sm}-\Sigma_{0\sm}(\omega) + \frac{U\Sigma_{2\sm}(\omega)}{\omega-\veps_{\sm}-U-\Sigma_{0\sm}(\omega)-\Sigma_{1\sm}(\omega)}}
+ \frac{\overline{\nbraket{n_{\bsm}}}}{\omega-\veps_{\sm}-U-\Sigma_{0\sm}(\omega) + \frac{U\left[\Sigma_{2\sm}(\omega)-\Sigma_{1\sm}(\omega)\right]}{\omega-\veps_{\sm}-\Sigma_{0\sm}(\omega)-\Sigma_{1\sm}(\omega)}}\,.
\label{eq:LacroixFU}
\edq
It is worth to consider the limit $U \to \infty$ where the impurity Green function expression \eqref{eq:LacroixFU} is greatly simplified as
\beq
\dbraket{d_{\sm}} = \frac{1 - \nbraket{n_{\bsm}} - \sum_{\alpha,k} \frac{V_{\alpha k\bsm}\nbraket{d_{\bsm}^{\dag}c_{\alpha k\bsm}}}{\omega-\veps_{\alpha k\bsm}+\veps_{\bsm}-\veps_{\sm}}}
{\omega-\veps_{\sm}-\Sigma_{0\sm}(\omega) - \sum_{\alpha,k}\sum_{\beta,q} \frac{V_{\beta q\bsm}V_{\alpha k\bsm}\calf_{\beta q;\alpha k}^{\bsm}}{\omega-\veps_{\alpha k\bsm}+\veps_{\bsm}-\veps_{\sm}}
+ \sum_{\alpha,k} \frac{V_{\alpha k\bsm}\nbraket{d_{\bsm}^{\dag}c_{\alpha k\bsm}}}{\omega-\veps_{\alpha k\bsm}+\veps_{\bsm}-\veps_{\sm}}\Sigma_{0\sm}(\omega)}\,.
\label{eq:LacroixIU}
\edq
\end{widetext}
Now we follow Ref.~[\onlinecite{PhysRevLett.66.3048}] to obtain a much simple truncated impurity Green function by  setting $\nbraket{d_{\sm}^{\dag}c_{k\sm}} = 0$ and $\calf_{qk}^{\sm} = \delta_{k,q}f(\veps_{k\sm})$ . By doing this, then the impurity Green function is given by
\beq
\dbraket{d_{\sm}} = \frac{1 - \nbraket{n_{\bsm}}}
{\omega-\veps_{\sm}-\Sigma_{0\sm}(\omega) - \Sigma_{2\sm}(\omega)}\,,
\label{eq:MeirIU}
\edq
where 
\begin{widetext}
\beq
\Sigma_{2\sm}(\omega) \approx \sum_{\alpha,k} \frac{|V_{\alpha k\bsm}|f(\veps_{\alpha k\bsm})}{\omega-\veps_{\alpha k\bsm}+\veps_{\bsm}-\veps_{\sm}+i0^+}
\approx  -\sum_{\alpha}\! \frac{\Gamma_{\alpha\bsm}}{\pi} \!\Biggr[i\pi f_{\alpha\bsm}(\omega \! +\! \veps_{\bsm}\! - \!\veps_{\sm})+ \ln \frac{\sqrt{(\pi/\beta)^2 + (\omega \!+\! \veps_{\bsm} \!-\! \veps_{\sm} \!-\!\mu_{\alpha\bsm})^2}}{D}\Biggr]\,.
\label{eq:Sigma2}
\edq
\end{widetext}
According to Ref.~[\onlinecite{PhysRevLett.91.127203}], 
in Eq.~\eqref{eq:MeirIU} the bare dot level $\veps_{\sm}$ in $\Sigma_{1\sm}(\omega)$ must be replaced by the renormalized one $\weps_{\sm}$ that it must be self-consistently found from the relation
\beq
\weps_{\sm} = \veps_{\sm} + \Re\left[\Sigma_{0\sm}(\weps_{\sm}) + \Sigma_{1\sm}(\weps_{\sm},\weps_{\up},\weps_{\down})\right]\,.
\label{eq:selfcons}
\edq
By doing this, we include spin-dependent high-order contributions that produce the effective exchange field when spin-dependent tunneling couplings are considered. 
\bibliography{spinbias}
\end{document}